\documentclass[twocolumn,showpacs,superscriptaddress,amsmath,amsfonts,amssymb]{revtex4-1}
\usepackage{units}
\usepackage{amssymb}
\usepackage{graphicx}
\usepackage{wasysym}

\begin{document}
\title{New perspectives in the ultrafast spectroscopy of many-body excitations in correlated materials}
\author{C. Giannetti}
\affiliation{Department of Physics \& i-Lamp, Universit\`a Cattolica del Sacro Cuore, Brescia I-25121, Italy}

\begin{abstract}
\textbf{Ultrafast spectroscopies constitute a fundamental tool to investigate the dynamics of non-equilibrium many-body states in correlated materials. Two-pulses (pump-probe) experiments have shed new light on the interplay between high-energy electronic excitations and the emerging low-energy properties, such as superconductivity and charge-order, in many interesting materials. Here we will review some recent results on copper oxides and we will propose the use of high-resolution multi-dimensional techniques to investigate the decoherence processes of optical excitations in these systems. This novel piece of information is expected to open a new route toward the understanding of the fundamental interactions that lead to the exotic electronic and magnetic properties of correlated materials.}     
\end{abstract}

\maketitle

\section{The interplay between the charge-transfer excitation and the low-energy phenomena in copper oxides}
\label{interplay}
\begin{figure*}
\includegraphics[width=13.5cm]{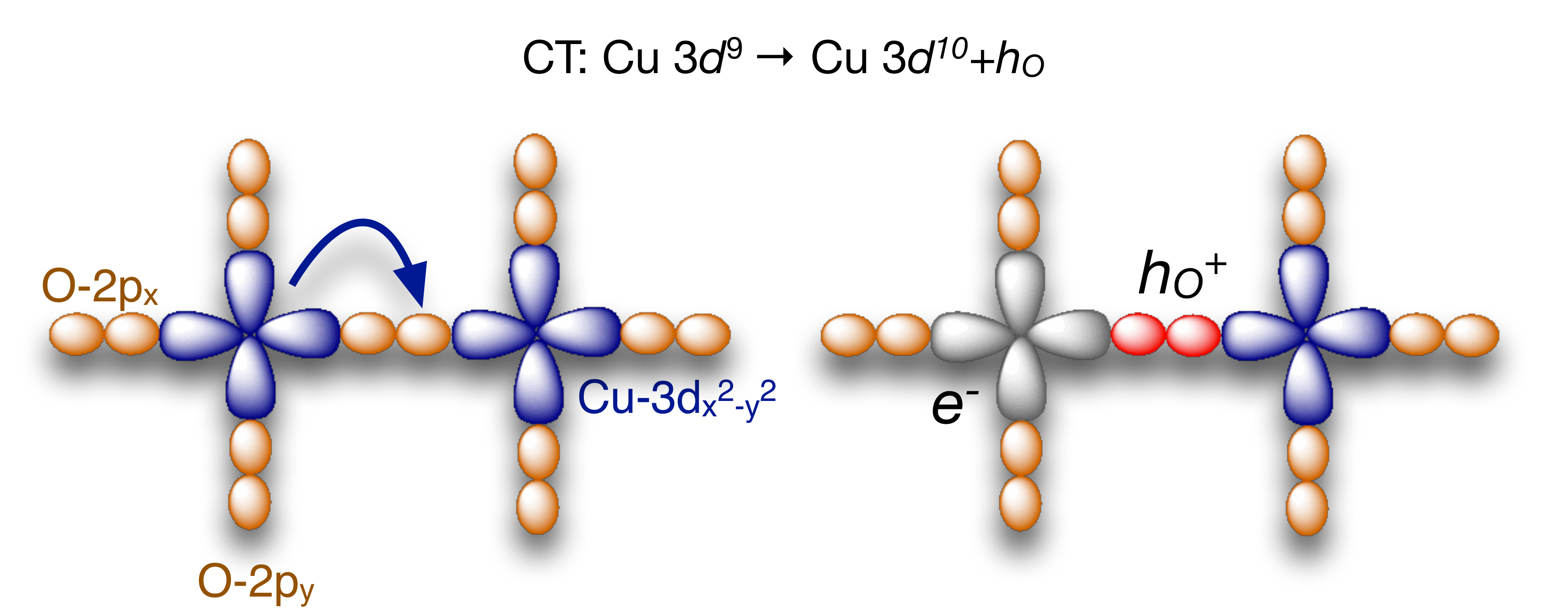}     
\caption{Sketch of the charge-transfer process in the Cu-O layer of copper oxides.}\label{fig_CT}
\end{figure*}

One of the most striking consequences of the strong correlations in materials is that extremely different energy scales, that are usually completely independent, become strongly intertwined up to the point that low-energy phenomena, such as superconductivity, strongly affect the electronic properties at energies of the order of several electronvolts.
One of the most celebrated examples is given by the Mott physics in correlation-driven insulators. When a small fraction of holes $x$ is doped into the insulator, the number of combinations available for having either a doubly- or un-occupied site is $N(1-x)$, that corresponds to a $Nx$ decrease of the spectral weight of both the Upper Hubbard Band (UHB) at +$U$/2 and of the Lower Hubbard Band at -$U$/2, $U$ being the onsite Coulomb repulsions. In order to conserve the total number of states, the variation of the high-energy spectral weight is compensated by the emergence of 2$xN$ states at the Fermi level, corresponding to the two ways (spin up and down) of occupying the $xN$ empty sites \cite{Meinders1993,Phillips2006}. As a consequence, the low-energy conducting states that host the strongly interacting quasiparticles, are created at the expenses of high-energy electronic states, in contrast to conventional materials in which the \textit{independent-electron approximation} renders the properties of low-energy quasiparticles completely independent from the high-energy physics.

\begin{figure}
\begin{centering}
\includegraphics[width=9cm]{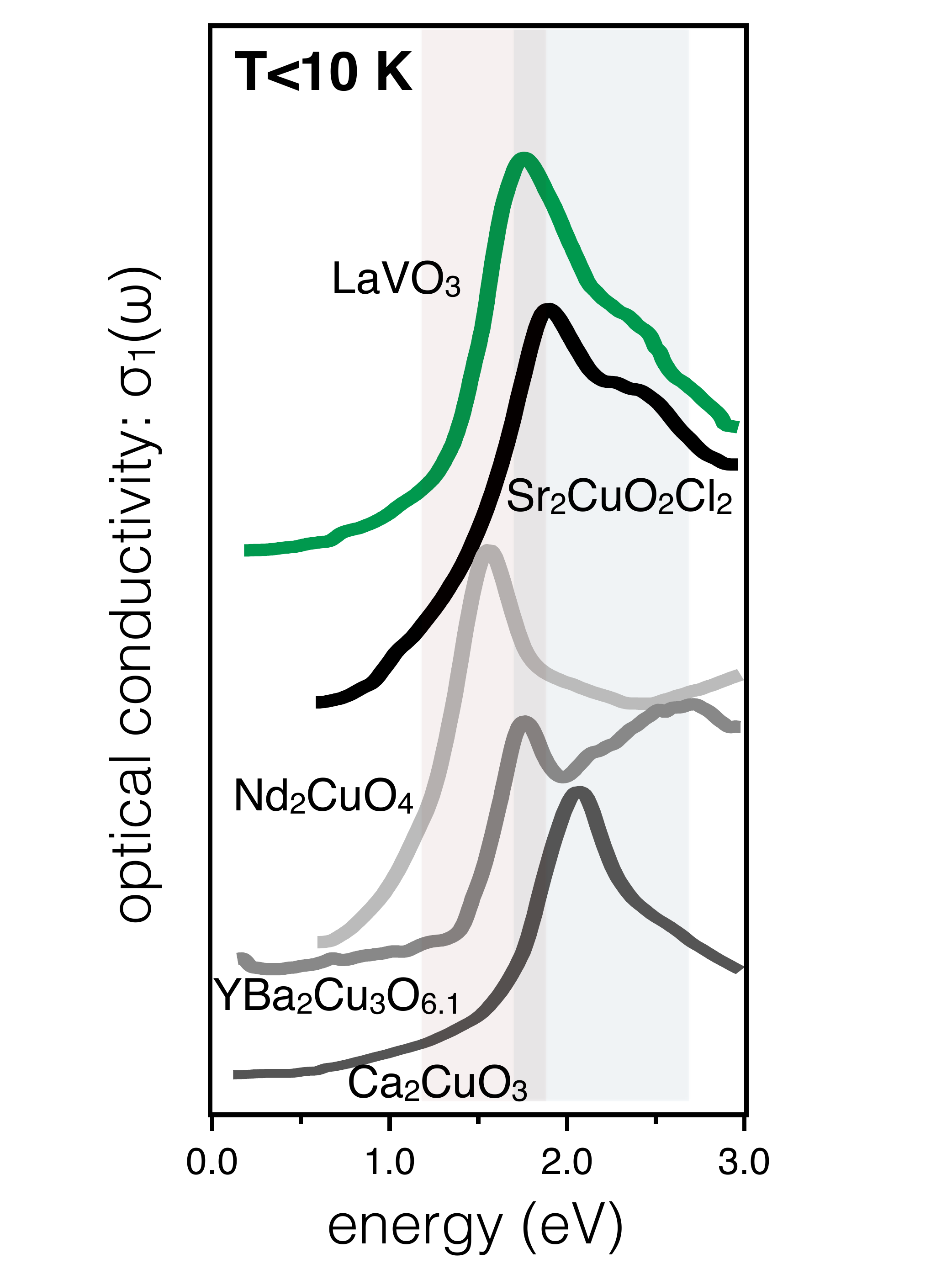}     
\caption{Low temperature ($T$=10 K) charge-transfer and exciton resonances for different families of transition metal oxides \cite{Uchida1991,Miyakasa2002}. The coloured areas represent the energy regions that can be accessed by sub-10 fs NOPA pumped by amplified ultrafast lasers.}\label{fig_CTedge}
\end{centering}
\end{figure}

Although much more complex, also the physics of charge-transfer insulators (CTI), such as copper oxides, is expected to violate the low-energy scale sum rule. In copper oxides the charge fluctuations within the Cu-3$d^9$ orbitals are strongly suppressed by the Coulomb repulsion ($U\sim10$ eV) between two electrons occupying the same Cu orbital. The lowest excitation is thus the charge-transfer (CT) of a localized Cu-3$d_{x^2-y^2}$ hole into its neighbouring O-2$p_{x,y}$ orbitals (see Fig. \ref{fig_CT}), with an energy cost, $\Delta_{CT}\sim$2 eV, smaller than the onsite Coulomb repulsion. This fundamental excitation is clearly observed by optical spectroscopy. As reported in Fig. \ref{fig_CTedge}, the optical conductivity of prototypical insulating transition-metal oxides shows an ubiquitous CT edge at $\hbar\omega$=$\Delta_{CT}$ \cite{Uchida1991,Miyakasa2002}, which defines the onset of optical absorption by particle-hole excitations in the complete absence of a Drude response. In principle, the information about the dynamics of the CT excitation is encoded into the CT structures visible at the edge of the absorption spectra reported in Fig. \ref{fig_CTedge}. These \textit{resonances} are mostly related to the formation of optical excitons, as a consequence of the local interaction between the photoexcited O-2$p_{x,y}$ holes and the neighboring Cu-3$d_{x^2-y^2}$ electrons. In this case, the Coulomb interaction tends to bind an electron and a hole giving rise to a resonance, the energy of which is below (or at the edge of) the excitation threshold of the unbound electron-hole pairs. Although the nature (i.e. symmetry, selection rule, dynamics) of these excitons has been subject of intense debate \cite{Miyakasa2002,Gossling2008}, the CT excitations are expected to provide a powerful link between the low- and high-energy physics in transition metal oxides. A simple example is given by the Zhang-Rice exciton, that has been observed as a polarization-sensitive shoulder at the CT edge in CuGeO$_3$, i.e. an edge-sharing CTI \cite{Pagliara2002}. The final state of this optical transition corresponds to the formation of a spin-singlet state on the neighboring plaquette, analogously to the formation of the low-energy Zhang-Rice singlet (ZRS) when a single hole is doped into the O-2$p_{x,y}$ bands of the insulating state \cite{ZhangRice1988}. In this example, the mechanism that drives the formation of the fundamental low-energy excitations, i.e. the ZRS, is mapped into the dynamics of high-energy CT excitons. 

When considering copper oxides, the interplay between the high- and low-energy scales leads to a variety of  manifestations that lack an explanation within the conventional band theory. As an example, a strong temperature dependence of the electronic bands over an energy range that is two orders of magnitude larger than the temperature scale involved has been observed by ARPES \cite{Kim2002}. This result demonstrates that, already in the normal state, the electronic correlations strongly affect the states at large binding energies.
On the other hand, continuous-wave optical techniques have revealed a superconductivity-induced variation of the optical properties at energies in excess of 1 eV \cite{Deutscher2005}. This effect, never measured in conventional superconductors, further supports the unconventionality of the superconducting mechanism in cuprates.\\

\section{Non-equilibrium approach}
Mapping the low-energy phenomena of correlated materials into the high-energy physics presents many advantages that have been exploited also by non-equilibrium approaches based on the use of ultrashort light pulses:
\begin{itemize}
\item The CT energy usually matches most of the  spectrum (1.5-3 eV) that can be covered by optical parametric amplifiers and other non-linear devices pumped by the output of Ti:sapphire ultrafast lasers \cite{Brida2010}.
\item The CT energy scale is much larger than the thermal energy $k_BT$. Therefore, the fingerprint of the electronic correlations in the ultrafast dynamics should be observed already at high temperatures. In this case, the ultrafast excitation can be used to selectively stimulate the CT charge fluctuations that remain freezed at the equilibrium temperature of the experiment. 
\item The CT excitation is preserved also when charge carriers are chemically doped in
the system. As a consequence, the CT dynamics can be used to track the evolution of the correlation effects as a function of doping. 
\end{itemize}

Taking inspiration from these considerations, ultrafast optical spectroscopies in the pump-probe configuration \cite{Giannetti2016} have been recently used to investigate the interplay between the physics of the CT and the emergence of ordered phases, such superconductivity and charge-order. In these experiments an ultrafast laser pulse (pump pulse) is used to perturb the ground state of the system, while a second delayed pulse monitors the evolution of the optical properties in the near-infrared/visible range back to the initial equilibrium state. Probing the photoinduced variation of the reflectivity (or optical conductivity) with a broad spectrum or with a tunable-wavelength pulse is crucial to perform a time-resolved optical \textit{spectroscopy} and to measure the temporal dynamics of the dielectric function on the femtosecond time scale. Since in any pump-probe scheme only the delay between the pump and the probe pulses is externally controlled, these techniques are often referred to as one-dimensional (1D) spectroscopies.

\subsection*{Charge-transfer dynamics and unconventional superconductivity}
The development of time- and frequency-resolved spectroscopies, based on the supercontinuum light produced by a photonic crystal fiber seeded by short infrared pulses, recently allowed to demonstrate that the dynamics of specific many-body electronic excitations at 1.5 and 2 eV, i.e., the energy scale of the CT process, is strongly affected by the onset of superconductivity in Bi-based double layer copper oxides (Bi2201) \cite{Giannetti2011}. This result allowed to finally solve the question about the origin of the superconductivity-induced modifications of the high-energy optical properties observed by equilibrium optical spectroscopies \cite{Deutscher2005,Molegraaf2002}. Furthermore, it was shown that the superconductivity-induced change at the CT energy strongly depends on the hole concentration of the system. In particular, the observed transient spectral weight change suggests a transition from a kinetic- to a potential-energy driven pairing mechanism \cite{Giannetti2011} at a relative doping of the order of 0.16, which correspond to the concentration necessary to attain the largest critical temperature of the systems ($T_c$=96 K). 

\subsection*{Charge-transfer dynamics and charge-order in cuprates}
The experimental evidence  of  charge  order  (CO)  in  cuprate  superconductors triggered a revival of the interest for the physics of copper oxides. CO appears as a universal phenomenon  in  hole-doped  systems  (evidences  for  CO  have  been  found  in  Y- \cite{Ghiringhelli2012,Chang2012,Achkar2012},  Bi- \cite{Comin2014,daSilva2014}  and  Hg \cite{Tabis2014} based  compounds, and  recently  also  in  electron-doped  materials,  such  as  NCCO \cite{daSilva2015}.  The  universality  of  this  phenomenon  raises  obvious  questions  about  the  mechanism  responsible  for  the  charge-ordering  tendency  in  cuprates  and  its  relation  with  the  superconducting  phase.  One  of  the  main  unsolved  questions  is  whether  the  CO  is  just  the  consequence  of  a  lattice-driven folding of the Fermi surface $-$ as in the conventional charge-density-wave instability $-$ or it is the consequence of a pure electronic mechanism in which the short-range correlations play a major role. 
\begin{figure*}
\includegraphics[width=22.5cm]{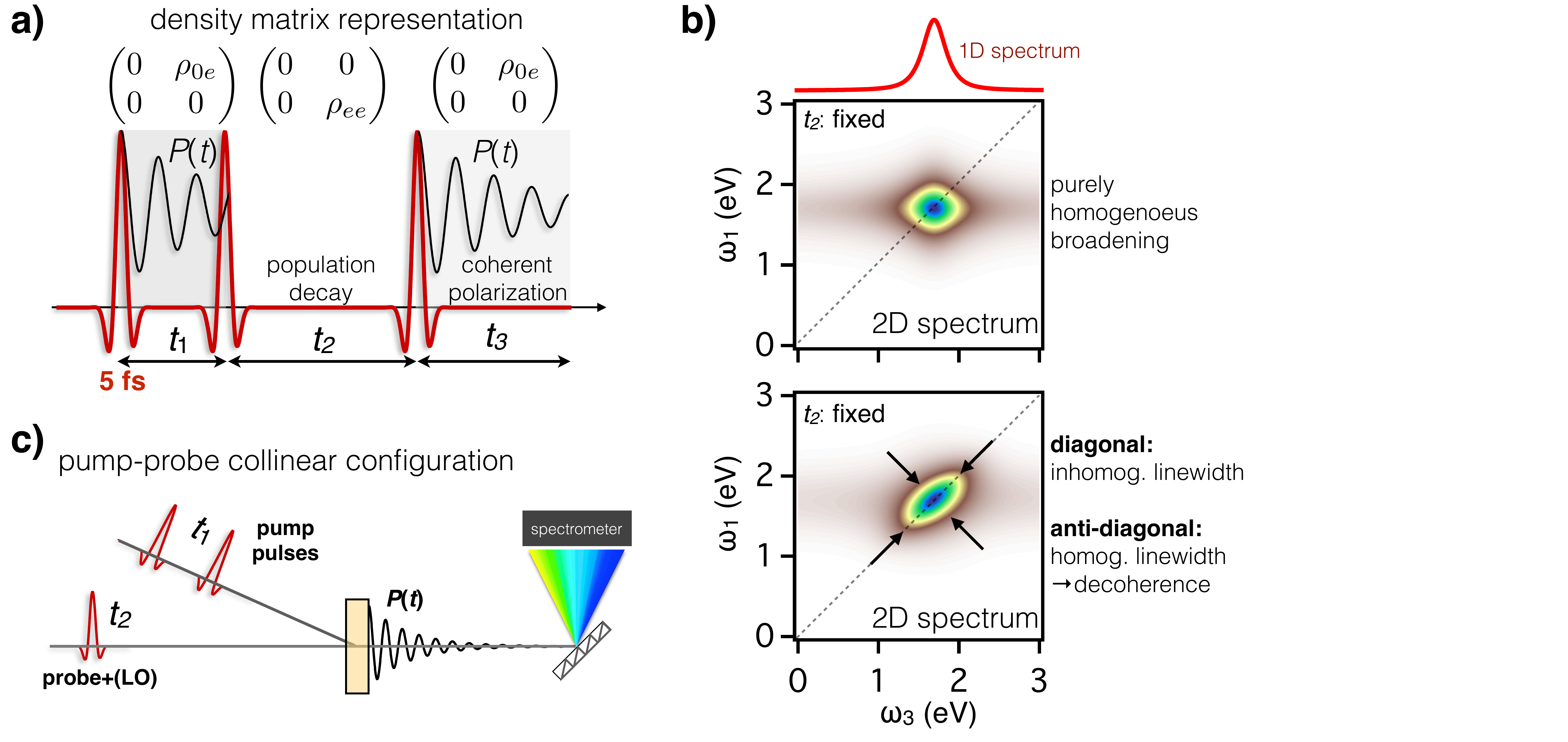}     
\caption{a) Cartoon of the 2D spectroscopy working principles. b) The panels report simulated 2D absorptive spectra for a 0.4 eV broad linewidth, representing the typical charge-transfer exciton in transition metal oxides. In the top panel the linewidth is entirely related to homogeneous broadening. In the bottom panel, the inhomogeneous broadening is the main mechanism that determines the total linewidth. The top red curve represents the corresponding information that would be provided by a broadband pump-probe experiment. The experimental resolution (5 fs) is taken into account. c) Partially collinear Pump-probe geometry for 2D spectroscopy. The probe pulse is self-heterodyned and is dispersed onto a spectrometer.}\label{fig_2D}
\end{figure*}

Time-resolved optical spectroscopy in La-substituted single layers cuprates (La-Bi2201) recently unveiled \cite{Peli2015} a sharp and drastic change  in  the  character  of  the  Cu-3$d\rightarrow$O-2$p$  CT process,  that  turns  from  a  localized  transition $-$ as  in  a  Mott  insulator $-$ to a delocalized process $-$ as in a conventional metal. This change is observed exactly at the doping at which  the low-temperature CO vanishes ($p_{cr}\sim$0.16), suggesting an intimate relation between the CO process and the CT dynamics. These results point to the existence of an underlying correlated state, that is  precursor to the low-temperature CO and, possibly, to the other instabilities that fan out from a zero-temperature quantum critical point at $p_{cr}$.

\section{Multi-dimensional spectroscopy to access the decoherence time}
Despite the recent results that link the ultrafast dynamics of the CT excitations to the low-energy properties of copper oxides, the most fundamental cause-effect relation and the role played by the high-energy excitations in the development of  symmetry-broken phases is still unclear. In this sense, the main limitation of any pump-probe scheme \cite{Giannetti2016} is that, as already discussed in Section 2, the two-pulses techniques simply probe the depopulation of a specific excited state, while they remain blind to the fundamental decoherence processes. This limitation is intrinsic to the 1D nature of pump-probe experiments, irrespective of the polarization and frequency (from THz to UV) of the light pulses employed.

Accessing the dynamics of the dephasing of the macroscopic polarization would unlock the gate to the study of the way the initial CT excitations loose their quantum coherence and transform into incoherent charge fluctuations. Since
the electronic decoherence is driven by the interaction with the fluctuations of the environment (e.g. charge-distribution, short range magnetic correlations, phonons),
the measurement of the electronic dephasing would greatly help in clarifying how the onset of long-ranged orders affects the decoherence processes and it would provide novel insights into the most relevant interactions that regulate the electron dynamics in correlated materials. 

Here we suggest a new route to overcome the main limitations of pump-probe techniques and to access the ultrafast dephasing of CT excitations. To this purpose we need a technique characterized by: i) a time resolution as short as the typical decoherence timescale ($<$10 femtoseconds) in correlated materials; ii) the capability of disentangling the real polarization dephasing from the population decay and from inhomogeneous broadening effects. These requirements can be fulfilled by multi-dimensional spectroscopy, i.e. the optical analogue of NMR techniques, that has been developed over the past two decades in physical chemistry and atomic physics \cite{Cundiff2013}. In particular, two-dimensional spectroscopy makes use of a suitable sequence of three coherent light pulses (see Fig. \ref{fig_2D}) to trigger an initial macroscopic polarization and to track its subsequent dephasing dynamics. In the simplest picture (see Fig. \ref{fig_2D}), the first pulse creates a macroscopic polarization, leaving the system in the coherent superposition between the ground, $\vert\Psi_0\rangle$, and the excited state, $\vert\Psi_e\rangle$. The polarization, described by the off-diagonal terms ($\rho_{0e}$) of the density matrix, oscillates in time with frequency $\omega_{0e}$ and decays on the dephasing timescale $\tau_D$. The second pulse, at time $t_1$, converts the macroscopic polarization into a population in the excited state, which does not oscillate but is reminiscent of the initial phase of the polarization. The third pulse, at time $t_2$, converts back to a coherent polarized state that radiates the measured signal field at time $t_3$. The two-dimensional spectrum (see the example in Fig. \ref{fig_2D}) is generated, for each fixed $t_2$, by scanning $t_1$ and $t_3$ and taking the Fourier transforms. In the last decades, many different schemes for mixing the two coherent pump pulses and the probe have been designed and used \cite{Hamm2011}. In the most general non-collinear case, the phase matching conditions enable to disentangle the "rephasing" and "non-rephasing" processes, which depend on the phase of the polarization created by the probe pulse with respect to that induced by the pump pulses. However, the simplest scheme for investigating the CT dynamics is the partially collinear pump-probe geometry \cite{Hamm2011}, which employs two phase-locked collinear pump pulses and a noncollinear probe pulse that is dispersed on a spectrometer ($\omega_3$), thus working in a self-heterodyning configuration (see Fig. \ref{fig_2D}c). The advantages of this configuration are its simplicity, its applicability to standard pump-probe systems, and the fact that it automatically measures absorptive 2D spectra, that are the most direct to interpret. Two-dimensional spectroscopy is considered the ultimate non-linear technique, since it contains all the information that can be retrieved by 0D (conventional optics) and 1D spectroscopies, such as multi-colour pump-probe and multi-pulse techniques without either the pump- or probe-frequency resolution \cite{Yusupov2010,DalConte2012,Madan2014,DalConte2015}. On the other hand, 2D spectroscopy also contains fundamental information that is absent in 0D and 1D, such as:
\begin{itemize}
\item the capability of discriminating between homogeneous and inhomogeneous linewidths \cite{Hamm2011}
\item the possibility of following the exciton migration and the charge-separation process \cite{Abramavicius2010}.
This technique has been already applied, for example, to investigate many-body excitonic effects in semiconductors at $T<$10 K \cite{Zhang2007}.
\end{itemize}

The dramatic advances in fiber-laser technology made recently available the laser sources necessary to bring the existing 2D spectroscopy schemes into the regime necessary to investigate the decoherence dynamics in correlated materials at low temperature. In particular, the key technological requirements are: i) the extreme temporal resolution ($\sim$5 femtoseconds), to access the relevant timescales on which the coupling with the thermal bath (phonons, orbital and spin fluctuations) occurs; ii) an unprecedentedly high repetition rate (up to the MHz), allowing to work in the low-fluence ($<$10 $\mu$J/cm$^2$) regime and investigating the low-temperature phases (superconductivity, charge-order, magnetic order) without melting them \cite{Giannetti2009,Giannetti2011}.
To achieve these performances it is possible to exploit the recently developed amplified fiber-lasers that can be used to pump Non-Collinear-Optical-Parametric-Amplifiers (NOPAs) in order to obtain few femtosecond pulses in the 1.2-1.9 eV and 1.7-2.6 eV energy ranges \cite{Nillon2014,Liebel2014}.  

The feasibility of the proposed experiment is crucially based on the possibility of finding well defined excitonic transitions in correlated oxides, whose dephasing time can be measured by 2D spectroscopy. As shown in Fig. \ref{fig_CTedge}, many of the CT excitonic lines, although strongly affected by inhomogeneous broadening, are entirely contained in the spectral band of the 2D spectroscopy based on amplified fiber-laser. This demonstrates the possibility of directly probing the fundamental decoherence of the CT excitation. The same conclusion can be drawn by analysing the problem in the time-domain: the decoherence of the CT excitons is mainly related
to the the coupling with spin and lattice fluctuations and to the
direct interaction with other excitations, which can be limited by
working at very low excitation fluence. In the worst scenario, quantum coherence is lost at each hopping process in the lattice, which is in the order of 2-6 fs if we assume a typical hopping integral of 0.1-0.3 eV for transition-metal oxides. However, there are many effects that concur to increase this value: i) the formation of "dressed"
coherent quasiparticles is expected to increase the effective mass, thus reducing their mobility within the lattice \cite{Kane1989,Lee2006}; ii) in multi-band systems, the possibility of hopping on the oxygen sub-lattice strongly reduces the effect of spin-fluctuations on the electron dynamics \cite{Ebrahimnejad2014}. The typical timescale of the decohence of CT excitations is thus fully accessible by the 2D spectroscopy that can be developed with the available technology.

To show the advantages of 2D spectroscopy with respect to broadband pump-probe techniques, a simulation of the absorptive 2D spectrum expected for a typical CT excitation in a copper oxide is reported in Fig. \ref{fig_2D}b. The simulation takes into account the expected temporal resolution of the experiment (5-6 fs). Assuming a total linewidth of 400 meV (the value corresponding to the excitonic line of Nd$_2$CuO$_4$), the inhomogeneously (bottom panel) and homogeneously (top panel) broadened cases are clearly disentangled in the 2D spectra. In contrast, any 1D (or 0D) optical spectroscopy would only access the integral with respect to $\omega_1$ of the 2D spectrum and would not be able to discriminate between the two cases. The 2D spectroscopy can thus be used to access the intrinsic decoherence time of CT excitations and to understand its interplay with the onset of macroscopic condensates, such as long-range antiferromagnetism, superconductivity and charge-order for copper oxides or orbital order for lanthanum vanadates.

\section{Perspectives: accessing the decoherence of many-body excitations in correlated materials}

\begin{figure*}
\includegraphics[width=17.5cm]{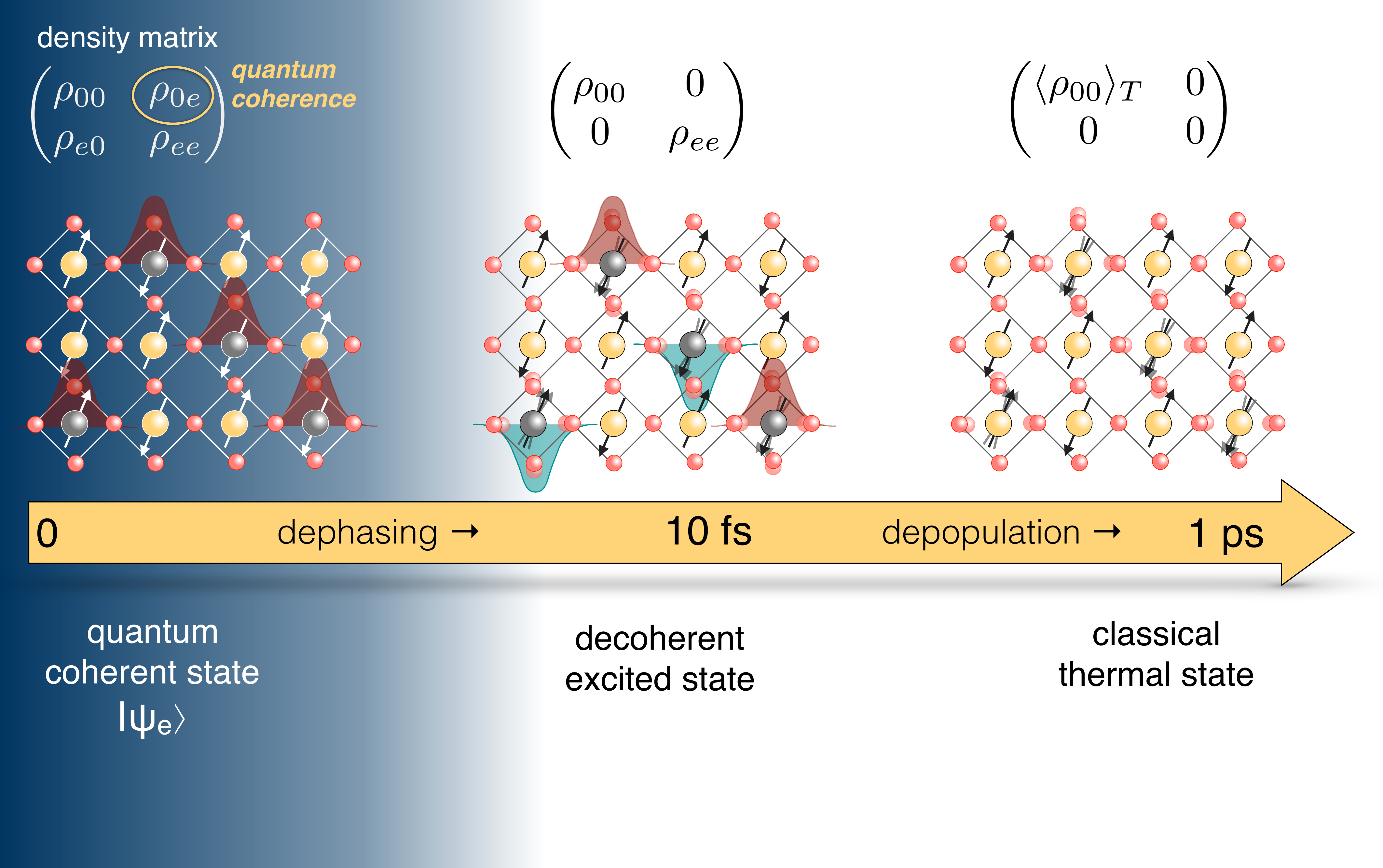}     
\caption{Cartoon of the evolution of coherent many-body excitations in correlated materials. Within few femtoseconds  (1-10 fs), the coherence, represented by the off-diagonal terms of the density matrix, is completely lost due the coupling with the thermal bath (e.g., phonons, magnetic excitations). On a longer timescale, the incoherent population of photo-injected excitations recombines due to inelastic process, leading to a classical thermal state at a larger effective temperature.}
\end{figure*}

Accessing the decoherence of optical excitations on the timescale relevant for the most important interactions among the electrons and with lattice vibrations or with other degrees of freedom (e.g. magnetic, orbital) is expected to open intriguing possibilities. As discussed in Sec. \ref{interplay}, the CT excitation is expected to be intrinsically intertwined with the low-energy degrees of freedom. As an example, let us consider the charge-transfer of a localized Cu-3$d_{x^2−y^2}$ hole to its neighbouring O-2$p_{x,y}$ orbitals in the insulating state, which forms a local exciton moving in the surrounding lattice. The transition from an ordered antiferromagnetic background to a strongly fluctuating spin background is expected to significantly affect the decoherence dynamics. Similarly, the CT coherent dynamics is expected to be strongly sensitive to fluctuations of the electronic density (charge order) and to the Cooper-pair condensation. In perspective, we can list some selected unsolved questions that could be tackled by the proposed multi-dimensional spectroscopy: 
\begin{itemize}
\item How do many-body excitations in quantum materials driven out of equilibrium dephase and eventually achieve equilibration? In the archetypal case of insulating copper oxides, what is the hierarchy of the interactions with the magnetic, orbital and lattice degrees of freedom in leading the dephasing and thermalization?
\item How does the vicinity to the Mott (charge-transfer) insulating phase and the intrinsic structural and electronic inhomogeneities affect the dephasing dynamics of the fundamental charge-transfer excitations in weakly-doped cuprates?
\item How does the emergence of low-temperature quantum macroscopic phases, such as superconductivity and charge-ordering, affects the intrinsic dephasing and thermalization? Is it possible to infer the nature of the fundamental interactions that drive superconductivity and charge-ordering?
\item Is it possible to directly access the coherent dynamics of the Mott (charge-transfer) gap before it is washed out by the dephasing processes? Is it possible to manipulate the electronic properties of materials on the verge of correlation-driven phase transition by creating suitable coherent superpositions of quantum states?
\item What is the origin of the intra-unit cell electronic nematicity that characterizes the pseudogap state of hole-doped cuprates and of the critical fluctuations in the vicinity of the quantum critical point at hole doping $\sim$0.16?
\item What is the nature of the local electronic and structural inhomogeneities in the vicinity of the Mott transition?
\end{itemize}
Furthermore, the proposed 2D spectroscopy could be combined with other excitation schemes based on mid-infrared and THz stimulation of specific phonon modes. Similar experiments could be implemented to exploit coherent X-ray pulses, generated by either free-electron lasers or table-top high-harmonic sources, to directly probe the fundamental dephasing processes of emerging excitations, such as excitons, polaritons, and polarons in bulk and hetero-structured transition metal oxides.\\

This paper was invited and supported by the Italian Physical Society (S.I.F).

\end{document}